\documentclass[journal=jacsat,manuscript=article]{achemso}
\DeclareUnicodeCharacter{0308}{} 

\usepackage[version=3]{mhchem} 
\usepackage{comment}
\usepackage{hyperref}
\usepackage{amsmath} 
\usepackage{svg}
\usepackage{xcolor}

\author{Edward Danquah Donkor}
\affiliation{The Abdus Salam International Center for Theoretical Physics, Strada Costiera 11, 34151
Trieste, Italy.}
\alsoaffiliation{SISSA –  via Bonomea 265, 34136 Trieste,
Italy.}
\author{Alessandro Laio}
\affiliation{SISSA – via Bonomea 265, 34136 Trieste,
Italy.}
\email{laio@sissa.it}
\author{Ali Hassanali}
\affiliation{The Abdus Salam International Center for Theoretical Physics, Strada Costiera 11, 34151
Trieste, Italy.}
\email{ahassana@ictp.it}

\title[An \textsf{achemso} demo]
  {
  Do Machine-Learning Atomic Descriptors and Order Parameters Tell the Same Story? The Case of Liquid Water.
  }

\abbreviations{IR,NMR,UV}

\begin{document}

\begin{abstract}
Machine-learning (ML) has become a key workhorse in molecular  simulations. Building an ML model in this context, involves encoding the information of chemical environments using local atomic descriptors. In this work, we focus on the Smooth Overlap of Atomic Positions (SOAP) and their application in studying the properties of liquid water both in the bulk and at the hydrophobic air-water interface. By using a statistical test aimed at assessing the relative information content of different distance measures defined on the same data space, we investigate if these descriptors provide the same information as some of the common order parameters that are used to characterize local water structure such as hydrogen bonding, density or tetrahedrality to name a few. Our analysis suggests that the ML description and the standard order parameters of local water structure are not equivalent. In particular, a combination of these order parameters probing local water environments can predict SOAP similarity only approximately, and viceversa, the environments that are similar according to SOAP are not necessarily similar according to the standard order parameters. We also elucidate the role of some of the metaparameters entering in the  SOAP definition in encoding chemical information.

\end{abstract}


\section{Introduction}

The last decade has seen a tremendous spurt in both the development and application of machine-learning (ML) approaches to study molecular systems\cite{behler2007generalized,montavon2013machine,faber2016machine}. ML has become a mainstream component of atomistic modeling. In particular, for the construction of  interaction potentials \cite{behler2007generalized,shapeev2016moment,smith2017ani},  and for the analysis of molecular dynamics  simulations\cite{glielmo2021unsupervised,matsunaga2018linking,botlani2018machine,liang2021supervised}. Establishing an understanding of the underlying physical and chemical principles that make ML useful, accurate and in short, avoiding its \emph{black box} usage, remains an open challenge.

A critical first step in designing a ML-based model for molecular systems, is identifying atomic descriptors that encode information of the local environments of a chemical moiety\cite{jinnouchi2020descriptors,low2020effect}. Over the recent years, various flavours of atomic descriptors have been developed including atomic-density ones which focus on characterizing \emph{local} environments\cite{behler2011atom,de2016comparing}, topology based descriptors relying on graph-theoretical approaches\cite{bilbrey2020look} to extract the connectivity patterns in data and many-body tensor network representations which serve to characterize the global structure of materials\cite{huo2017unified}, to name a few. Most of these descriptors require human intervention in the selection of various parameters. For example, the size of the local environment or the number of basis functions one needs to describe the local configurations. Since ML-based descriptors are designed to capture the physical and chemical nature of molecular systems, interpreting and understanding their meaning is important, and can lead to a more rational and also physical choice of the hyper-parameters entering their definition.

In this work, we address the issue of interpretability, focusing on a specific class of local-atomic descriptors namely; the Smooth Overlap of Atomic Positions (SOAP)\cite{de2016comparing,bartok2013representing}. SOAP has become widely popular for identifying structural finger prints in systems such as liquid water\cite{monserrat2020liquid,capelli2022ephemeral,offei2022high} and inorganic crystals\cite{bartok2018machine,reinhardt2020predicting} and more recently, to develop coarse-grained intermolecular potentials\cite{byggmastar2022multiscale}. Here we examine how SOAP fingerprints are related to standard order parameters (Figure \ref{opener}, right panel) that are typically used to describe the local structure of liquid water\cite{gallo2016water}. Moreover, we study how this relationship changes from bulk liquid water to the air-water interface; a prototypical system used to study the environmental effects of hydrophobic interactions\cite{cheung2019air,mauri2014structure,d2019protein}.

To measure the relationship between SOAP-based descriptors and standard order parameters, we deploy a recently developed  technique, dubbed as the Information Imbalance\cite{glielmo2022ranking} (IB). IB is a statistical test that determines the relative information content between different distance measures defined on the same data space. If applied to two different distances, it allows determining if these distances are equivalent, unrelated, or if one of the two is more predictive than the other.  Specifically, we explore the IB between SOAP descriptors and a wide variety of order parameters\cite{tet2,chau1998new,shiratani1996growth,saika2000computer,luzar1996effect,luzar1996hydrogen} that have been built on physical and chemical intuition to characterize aqueous environments.

Using the IB, we first investigate if a suitable combination of the order parameters is able to predict the similarity of local environments measured using SOAP features. We find that even the best combination of order parameters is able to predict the SOAP similarity only approximately. The quality of the prediction is better for configurations close to the surface of water, and is significantly improved by choosing a SOAP length scale parameter $\sigma = 0.25$ \AA, which is much smaller than the value typically used ($\sigma = 1$ \AA).  This result may not be too surprising as the structural information embedded in SOAP descriptors likely contains a richer characterisation of the local environments in water than the standard order parameters which are often fine-tuned to capture specific chemical interactions. 

We also investigate the reverse problem, namely if SOAP descriptors are able to predict the similarity as measured using chemical-intuition based order parameters. Rather surprisingly, we find that most of these standard order parameters can be predicted rather approximately. For example, the IB between the SOAP similarity and the similarity measured by the number of hydrogen bonds or the Local Structural Index (LSI), appears to be close to the value observed for distances which are unrelated to each other. Interestingly, SOAP predicts the standard coordination number the most accurately out of all the order parameters we examine. The IB can be improved rather marginally by reducing the value of $\sigma$. We provide some perspectives on the possible origins of these discrepancies.


The paper is organized as follows. We begin in \hyperref[methods]{Section 1}
with the Methods employed in this work, including a summary of both SOAP and order parameters we study, as well as the Information Imbalance technique. We then move on in \hyperref[results]{Section 2} to the Results where we illustrate the relationships we unravel between the SOAP and order parameters that are obtained using the IB method for both bulk and interfacial water. We then end in \hyperref[conclusion]{Section 3} with some conclusions of our work.

\section{Methods} \label{methods}

\subsection{Molecular Dynamics Simulations} 

Molecular Dynamics (MD) simulations are carried out using the LAMMPS package. \cite{lammpsref} We use the TIP4P/2005 \cite{abascal2005general} rigid water model for most of our work. We also repeat some of our analysis using the TIP3P\cite{jorgensen1983comparison} water model which is commonly used in bio-molecular simulations.

Our initial simulation setup consists of a bulk water system with 729 water molecules equilibrated at ambient temperature and pressure in a box with sides 27.9$\times$ 27.9$\times$ 27.9 \AA. \ To construct an interface, we add a vaccuum region of 139.5 \AA \ in the z-direction and then equilibrate within the NVT ensemble for 10 ns at 300 K with a timestep of 2 fs. This is followed by a production run of 20 ns. The velocity-rescaling \cite{bussi2007canonical} thermostat is used with a time constant of 100 fs. In our simulations, the real space cut-off for the Coloumb and Lennard Jones (LJ) interactions is 15 \AA. Long range corrections are treated using the Particle-Particle Particle-Mesh (PPPM)\cite{hockney2021computer}  solver for both the Coulomb and LJ interactions. In order to validate the use of our model for the air-water interface, we computed the surface tension for our simulation, obtaining a value of (67.98 $\pm$ \ 0.74)mN/m, consistent with previous reports.
\cite{vega2007surface}. For the TIP3P model, we obtain a surface tension of (47.29 $\pm$ 0.40) mN/m, which is also consistent with previous studies. \cite{vega2007surface}

\subsection{Identifying Bulk vs Surface Water Environments}

Using our MD simulations, a total of approximately 5000 local environments are sampled from the water-surface system, on which the SOAP power
spectrum is computed using the DScribe software package\cite{dscribe}. These environments are chosen by randomly selecting a water molecule every 4 ps. Since we were interested in understanding how the the relationship between SOAP fingerprints and order parameters evolves from bulk water to the hydrophobic surface of water, we characterised the interface using the Willard-Chandler Interface (WCI)\cite{willard2010instantaneous}. In brief, a coarse-grained density field is defined as a sum of Gaussian functions, with a specified smoothening parameter ($\xi$) centered on all the atoms. The interface is then chosen as the set of points for which the coarse-grained density is half of the bulk density. The $\xi$ value used for the WCI construction was 2.4 \AA \ consistent with previous studies. The water density as a function of distance from the WCI is then built, yielding the distribution shown in the Supporting Information (SI Figure 1). 

For our analysis, we define
various layers from the density profile as performed in several prior works\cite{pezzotti20172d,kessler2015structure}, which allows for identifying surface and bulk water molecules. In most of our analysis, we focus on comparing the bulk and surface as defined by those waters in layer 4 (Bulk) and layer 1 (L1) respectively (see SI Figure 1 for a visual depiction of these layers relative to the WCI).

\subsection{Descriptors for the Local Structure of Water}

\begin{figure}[!ht]
    \centering
    \includegraphics[width=15cm]{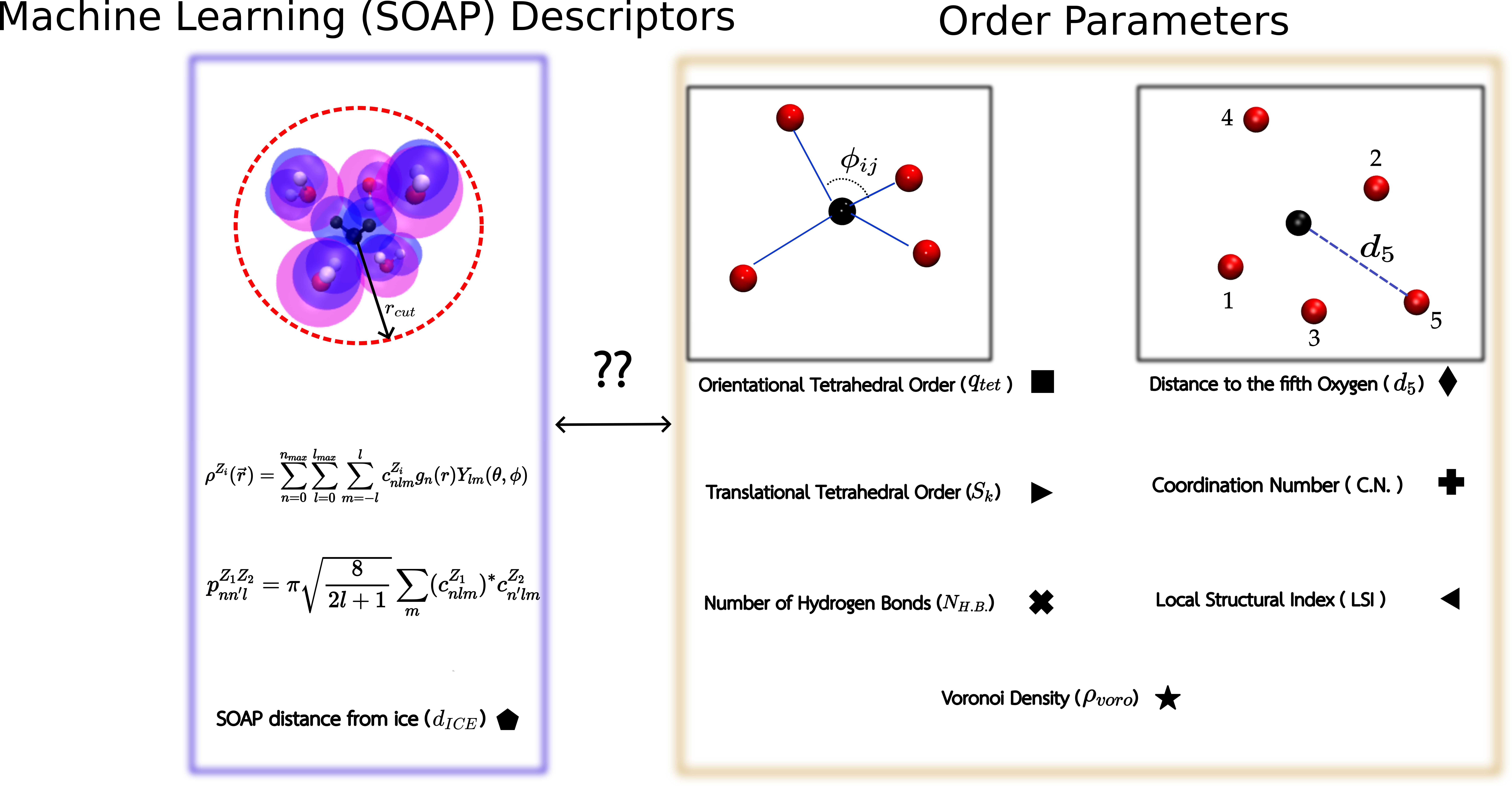}
    \caption{SOAP descriptors and order parameters (left and right respectively) used in this work. The left panel (top) shows a cartoon of the orientational and radial extent of the SOAP descriptors and the bottom panel summarizes the definition of the SOAP power spectrum - see main text for more details. The right panel shows snapshots of local water environments and some of the species, distance and angular based criteria that goes into defining the various order parameters. The symbols next to each variable is what is used throughout the manuscript to refer to each variable.}
        \label{opener}

\end{figure}

In this work, we examine the relationship between chemically inspired order parameters and SOAP-based descriptors using the Information Imbalance method. In the following, we begin by summarizing the theory underlying the construction of SOAP descriptors, (see \hyperref[soapsubsec]{Section 1.3.1}) and subsequently, the collection of different order parameters that we examine (see \hyperref[chemsubsec]{Section 1.3.2}). Finally, we also discuss the principles behind the Information Imbalance method (see \hyperref[chemsubsec]{Section 1.4}). Figure \ref{opener} illustrates the SOAP and order parameters along with the respective symbols that are used throughout the manuscript. 

\subsubsection{Smooth Overlap of Atomic Positions (SOAP)}
\label{soapsubsec}

SOAP has emerged in the last few years as a powerful method to describe the local environments of atoms and molecules, allowing for a wide range of applications from the study of structural properties of organic molecules\cite{maksimov2021conformational, grant2020network,de2016comparing} and very recently, the properties of liquid water\cite{offei2022high,capelli2022ephemeral,monserrat2020liquid}.

 In the context of SOAP, the local density of an atomic environment $\chi$ is written as a sum of Gaussian functions with variance $\sigma^2$, centered on all species that are neighbours of the central atom:
\begin{equation}
    \rho_{\chi} (\vec{r}) = \sum_{i \in \chi} \exp{\left(   \frac{-|\vec{r_i} - \vec{r}|^2}{2\sigma^2}\right)}
    \label{dens_ll}
\end{equation}

In this work, we will show that the choice of the value of $\sigma$ parameter plays a very important role in the ability of SOAP to predict chemical properties. Specifically, this parameter determines the resolution of chemical details of the water hydrogen bond network. 

The default value of $\sigma$ in the DScribe package is 1.0 \AA \ which to the best of our knowledge, is the value used in previous studies using SOAP to study the structure of liquid-water \cite{capelli2022ephemeral,monserrat2020liquid}. 

The atomic neighbour density in equation \ref{dens_ll}  can be expanded on a basis of radial basis functions and real spherical harmonics such that:
\begin{equation}
    \rho_{\chi}(\vec{r}) = \sum_{n = 0}^{nmax} \sum_{l = 0}^{lmax} \sum_{m = -l}^{l} c_{nlm} g_n (r) Y_{lm} (\theta, \phi)
\end{equation}

For practical purposes, one defines the environment $\chi$ by a cut-off radius ($r_{cut}$) and also limits the number of radial and angular basis functions ($n_{max}$, $l_{max}$) used. In the following work, we have examined the sensitivity of our results to changing $r_{cut}$ from 3.7 to 5.5 \AA \ as well as varying $n_{max}$ and $l_{max}$ between 10-12 and 6-8 respectively (see SI Figures 2-5). We find a marginal drop of chemical information contained in the SOAP descriptors (as reflected in the increase in the IB between SOAP and the order parameters) as we increase $r_{cut}$, owing to the fact that the order parameters we have examined, consider fluctuations only within the first solvation shell of a given water environment. We also find that there is no significant dependence of our results to the choice of $n_{max}$ and $l_{max}$. Unless otherwise stated, the $r_{cut}$, $n_{max}$ and $l_{max}$ are 3.7 \AA, 10 and 6 respectively.

By accumulating the expansion coefficients, a rotationally invariant power spectrum can be defined such that,

\begin{equation}
\label{ps}
    p_{nn'l} (\chi) = \pi \sqrt{\frac{8}{2l+1}}\sum_m (c_{nlm})^{\dagger} c_{n^{'}lm}
\end{equation}
Equation \ref{ps}  defines the components of the SOAP features we will use in our analysis.

\subsubsection{Order parameters} 
\label{chemsubsec}

\noindent \textit{Orientational Tetrahedral Order (q$_{tet}$)}:  

The $q_{tet}$ order parameter is one of the most used local structural quantities to describe liquid water\cite{chau1998new,tet1,tet2,tet3,tet4,tet5}. It measures how much a reference water environment deviates from an ordered tetrahedron whose vertices are defined by the bond vectors between the four nearest neighbouring oxygen atoms of a central one. It takes as input the angles between the O-O bond vectors and gives a value of 1 for a perfectly tetrahedral environment and closer to zero for environments which are not tetrahedral. Specifically, the tetrahedrality of a reference water molecule is defined as:

\begin{equation}
q_{tet} = 1 - \frac{3}{8} \sum_{i=1}^{3} \sum_{j=i+1}^{4} \left( \cos \phi_{ij} + \frac{1}{3}\right)^2
\end{equation} 

Where $\phi_{ij}$ is the angle between the oxygen molecule of the reference water and its nearest four neighbours (indexed with $i$ and $j$)

\noindent \textit{Translational Tetrahedral Order ($S_k$)}:

The translational tetrahedral order parameter is another measure of how much a reference water environment deviates from a regular tetrahedron\cite{chau1998new}. Whereas $q_{tet}$ focuses on the angles between the O-O bond vectors, $S_k$ is computed as the variance between the O-O distances of the four nearest water molecules to a central water:

\begin{equation}
    S_k = 1 - \frac{1}{3}\sum_{k = 1}^4 \frac{\left(r_k  - \bar{r}\right)^2}{4\bar{r}^2}
\end{equation}
$r_k$ is the distance between a reference water molecule and its $kth$ neighbour and $\bar{r}$ is the mean of these distances. It has been shown in previous works that $S_k$ is more sensitive to the local density variations compared to $q_{tet}$\cite{jedlovszky1999voronoi}.

\noindent \textit{Local Structural Index (LSI)}: 

The LSI is another important variable which has been used to study the structure of water in the bulk under different thermodynamic conditions\cite{appignanesi2009evidence,shiratani1996growth,malaspina2010structural}. It is obtained by ranking the O-O distances from an $ith$ central water molecule such that $r_1 < r_2 < ... < r_{i} < r_{i+1} < ... r_n < 3.7$ \AA $< r_{n+1}$, and then subsequently, the LSI is computed as:

\begin{equation}
\label{lsi} 
    \text{LSI} = \frac{1}{n}\sum_{i = 1}^n \left( \Delta(i) - \bar{\Delta}\right)^2
\end{equation}
where $\Delta(i) = r_{i+1} - r_i$ and $\bar{\Delta}$ is the arithmetic mean of $\Delta (i)$. A high value of LSI implies a larger separation between the first and second solvation shell and points to a more ordered water environment, while a lower value is interpreted as a more disordered environment.

\noindent \textit{Distance to the fifth Oxygen ($d_5$)}:

The $d_5$ is defined as the distance between a central oxygen atom and its fifth closest oxygen atom\cite{saika2000computer,cuthbertson2011mixturelike,tanaka2019revealing}. A large value of $d_5$ points to a high separation between the first and second solvation shell of a water environment and is interpreted as a locally ordered structure. A smaller value indicates a smaller separation between the first and second solvation shells and with a similar logic, a disordered local environment.

\noindent \textit{Coordination Number (C.N.)}: 

The coordination number quantifies the average number of atoms that surround a chosen central site within some
radial cutoff. It can be computed from the radial integral of the radial distribution function (RDF). In order
to have a smooth and continuous definition of the coordination number, we use a switching function, commonly done in the construction of different types of collective variables\cite{nelson1996namd,bonomi2009plumed}:

\begin{equation}
\label{switch}
    C.N. = \sum_{j = 1}^{N} \frac{1-\left(\frac{r_j}{r_{cut}}\right)^{12}}{1-\left(\frac{r_j}{r_{cut}}\right)^{28}}
\end{equation}
where $r_j$ is the distance between a central oxygen atom and atom $j$.

\noindent \textit{Number of Hydrogen Bonds ($N_{H.B.}$)}: 

We adopt the definition of hydrogen bonding by Luzar and Chandler\cite{luzar1996effect,luzar1996hydrogen}. This definition is based on a geometrical criterion and considers two water molecules to be hydrogen bonded when the distance between the donating and accepting oxygen atoms (O$_D$ and O$_A$) is within 3.5 \AA \ and the angle formed by the bond vector between the donating hydrogen and oxygen (H$_D$ and O$_D$) and the bond vector between O$_D$ and O$_A$ is less than 30$^\circ$. 

\noindent \textit{Voronoi Density ($\rho_{voro}$)}:

A quantitative measure of the local density ($\rho_{voro}$) in water can be extracted using a Voronoi
tessellation of the water network \cite{bernal1959geometrical,yeh1999orientational}. The $\rho_{voro}$ is then defined as the inverse of the Voronoi volume associated with a single water molecule which is a sum of the
atomic contributions coming from the oxygen and two hydrogen atoms. The Voronoi tesselation is carried out using the Voro$++$ code \cite{rycroft2009voro++}.

\noindent \textit{SOAP distance from ice (d$_{\text{ice}}$)}:

From the SOAP descriptors (power spectrum) show in equation \ref{ps}, it is then possible to define distances between structures. Several groups including ours have examined how SOAP environments in liquid water compare to different phases of ice\cite{capelli2022ephemeral,offei2022high,monserrat2020liquid}. In this work, one of these variables we use is comparing SOAP environments in liquid water to those in hexagonal ice (ice 1h) which is referred to as $d_{ice}$.
\begin{equation}
    d_{ice} = \sqrt{1 - \frac{\vec{p}(\chi_{\text{water}}) \cdot \vec{p}(\chi_{\text{ice 1h}}) }{|\vec{p}(\chi_{\text{water}})| |\vec{p}(\chi_{\text{ice 1h}})|}}
\end{equation}
Where $\vec{p}(\chi_{\text{water}})$ and $\vec{p}(\chi_{\text{ice 1h}})$ are the SOAP feature vectors for the liquid water and hexagonal ice environments respectively.
\subsection{Information Imbalance (IB)}
\label{ib}

The Information Imbalance (IB) is a recently developed method which can be used to quantify the relative amount of information between different types of variables which may or may not have the same measures of distance. The reader is referred to the original work for details\cite{glielmo2022ranking}. Here, we summarize the key ideas behind the method and why it provides a powerful tool for application in the context of the problems we address here.

Given a dataset with $N$ data points and characterised by $F$ features, we can define distance measures A and B  between the data points, such that A and B are computed using any subset of the feature space $F$ of choice.  With these distances in hand, the IB is then defined as:

\begin{equation}
    \Delta \left( A \rightarrow B \right) = \frac{2}{N} \langle R^{B} | R^{A }  = 1\rangle = \frac{2}{N^2}\sum_{i,j: R_{ij}^A = 1}R_{ij}^B
\end{equation}
Where $R_{ij}^{A}$ and $R_{ij}^{B}$ are the rank matrices obtained from distances A and B respectively, such that $R_{ij}^{A} = 1$ if $j$ is the first nearest neighbour of $i$ and $R_{ij}^{A} = 2$ if $j$ is the second nearest neighbour of $i$ and so on. With this definition, if $\Delta \left( A \rightarrow B \right) \sim 0$ this means A can fully predict B, while $\Delta \left( A \rightarrow B \right) \sim 1$ implies that A cannot predict B, as the ranks estimated with B are uncorrelated to those estimated with A. It is important to note that the IB is by definition asymmetric and thus one can examine predictability between different distances in both directions; if for example, $\Delta (A \rightarrow B) \sim 0.1$ and $\Delta (B \rightarrow A) \sim 0.4$, A will be able to predict B with better reliability than the reverse.

\section{Results}
\label{results}

\subsection{Predicting SOAP from Order Parameters}

In the ensuing analysis, we explore the relative information content between SOAP and order parameters using the IB. We begin by determining the ability of the order parameters to predict the full SOAP feature space and specifically, how sensitive this predictability is to the choice of the parameter $\sigma$ used in SOAP. Furthermore, we also examine how the IB changes as one moves from bulk water to the air-water interface.

\begin{figure}[!ht]
    \centering
    \includegraphics[width=15cm]{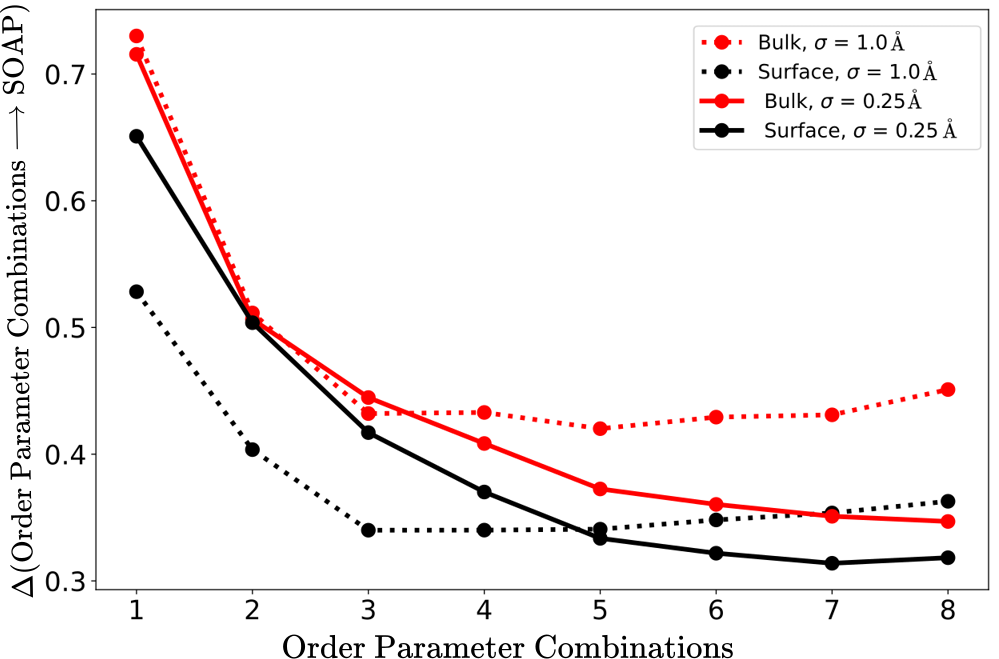}
    \caption{Information Imbalance between combinations of order parameters and the SOAP space. Comparison for Bulk (Red) and Surface (Black) and for $\sigma = 1.0$ \AA \ (Dotted line ) and $\sigma = 0.25$ \AA \ (Full line).  For $\sigma = 1.0$ \AA \, one needs a minimum of about 3 order parameters to describe the local fluctuations of the water environments contained in the full SOAP space, both in the bulk and at the Surface. For $\sigma = 0.25$, the number of order parameters needed increases by approximately 3.}
    \label{chem_soap}
\end{figure}

In Figure \ref{chem_soap} we show the IB between order parameters and SOAP, which encodes how many different order parameters are needed to predict the distances computed with the full high-dimensional SOAP vector. In order to obtain the IB for different combinations, we employ a greedy optimization whereby one iteratively selects the best combination of order parameters to minimize the IB between that combination and the SOAP space. This protocol can also be employed to select the $d$-dimensional compact SOAP descriptor that minimizes the Information Imbalance between SOAP and a target variable. For more details, we refer the reader to the original manuscript on the IB method\cite{glielmo2022ranking}.

Figure \ref{chem_soap}  
shows that using $\sigma = 1.0$ \AA, for both bulk and surface, a combination of three order parameters minimizes the IB to approximately 0.45 and 0.35 respectively. Recall, that an IB of 0.45 implies that if the data space included, say, 1000 local environments, the first neighbour environments according to the order parameters will be, on average the 225th neighbour according to SOAP. This is rather far from optimal since it suggests that local water environments predicted to be similar in the space of the order parameters, are quite distant using SOAP.

Using $\sigma = 0.25$ \AA, we notice a consistent increase in the number of order parameters needed to obtain the minimum IB; five order parameters are needed to obtain roughly the same IB as with $\sigma = 1.0$ \AA. This shows that by reducing the coarse-graining length of the Gaussian density from which the SOAP power spectrum was built, we essentially add more chemical information and consequently, this requires more order parameters to correctly describe. The sensitivity of the interplay between the size of the basis set and the choice of $\sigma$ in describing atomic environments has recently been discussed by Pozdnyakov and co-workers\cite{pozdnyakov2021local}. They show that in order to obtain the same sensitivity of the density expansion coefficients to vibrational distortions in a single methane molecule, the $\sigma$ should be approximately half that of the minimal interatomic distance.

\begin{figure}[!ht]
    \centering
    \includegraphics[width=15cm]{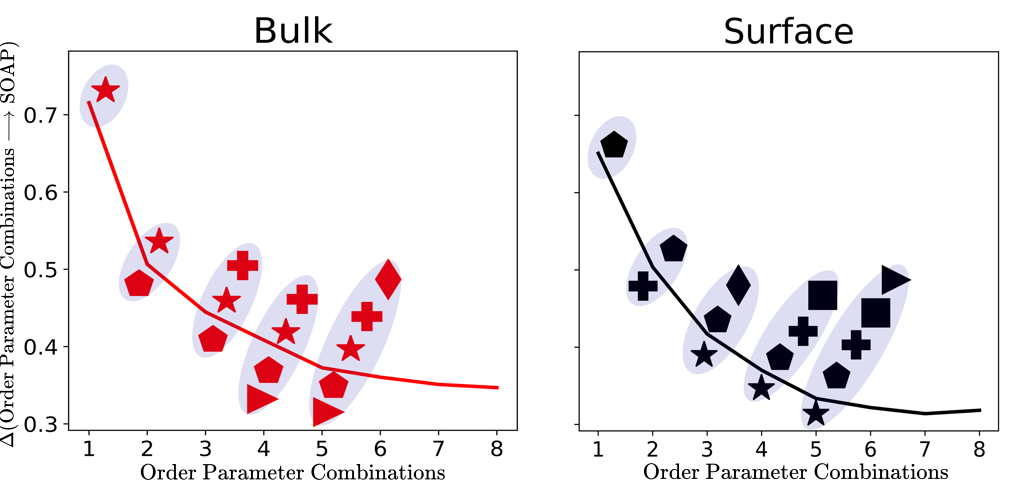}
    \caption{Information Imbalance between combinations of order parameters and the full SOAP space for $\sigma  = 0.25$, showing the selected variables for both bulk and surface. It is noteworthy that the difference between the Bulk and Surface for $\sigma = 0.25$ \AA \ is reflected in the difference between the combinations of variables needed to minimize the IB between the order parameters and the SOAP space. We can also see that $\rho_{voro}$ and $d_{ice}$ are consistently important to describe the SOAP space for both the bulk and surface whereas $q_{tet}$ for example becomes more important for the surface due to the enhanced orientational correlations at the Surface compared to the Bulk.}
    \label{select_chem_soap}
\end{figure}

While Figure \ref{chem_soap} shows that with $\sigma = 0.25 \ \text{\AA}$, the gap in the IB between the bulk and surface becomes much less pronounced compared to $\sigma = 1.0 \ \text{\AA}$, the combination of order parameters that are needed to describe the fluctuations in the bulk and the surface are not the same. In Figure \ref{select_chem_soap}, we show the order parameters that lead to the most predictive distance in the SOAP space. $\rho_{voro}$ and the one-dimensional SOAP based descriptor ($d_{ice}$) consistently appear to be important for both the bulk and surface. Remarkably, the LSI parameter which is often used to study local-water structure in the bulk\cite{poli2020charge,shiratani1996growth}, does not appear in any of the chemical combinations that lowers the IB. Interestingly, the tetrahedral order parameter, $q_{tet}$ plays a more important role at the surface due to the fact that there is an enhanced orientational order. These trends are consistently reproduced for simulations of the surface of water using the TIP3P water model (see SI Figure 8).

\subsection{Predicting Order Parameters from SOAP}

The preceding analysis examined how well the distances in the order parameter space can predict the full SOAP distances. Next we move on to discover the ability of SOAP to capture the information contained in single order parameters and at the same time, we use the IB to identify the optimal number of SOAP components required to perform a prediction. 

\begin{figure}[htp]
    \centering
    \includegraphics[width=15cm]{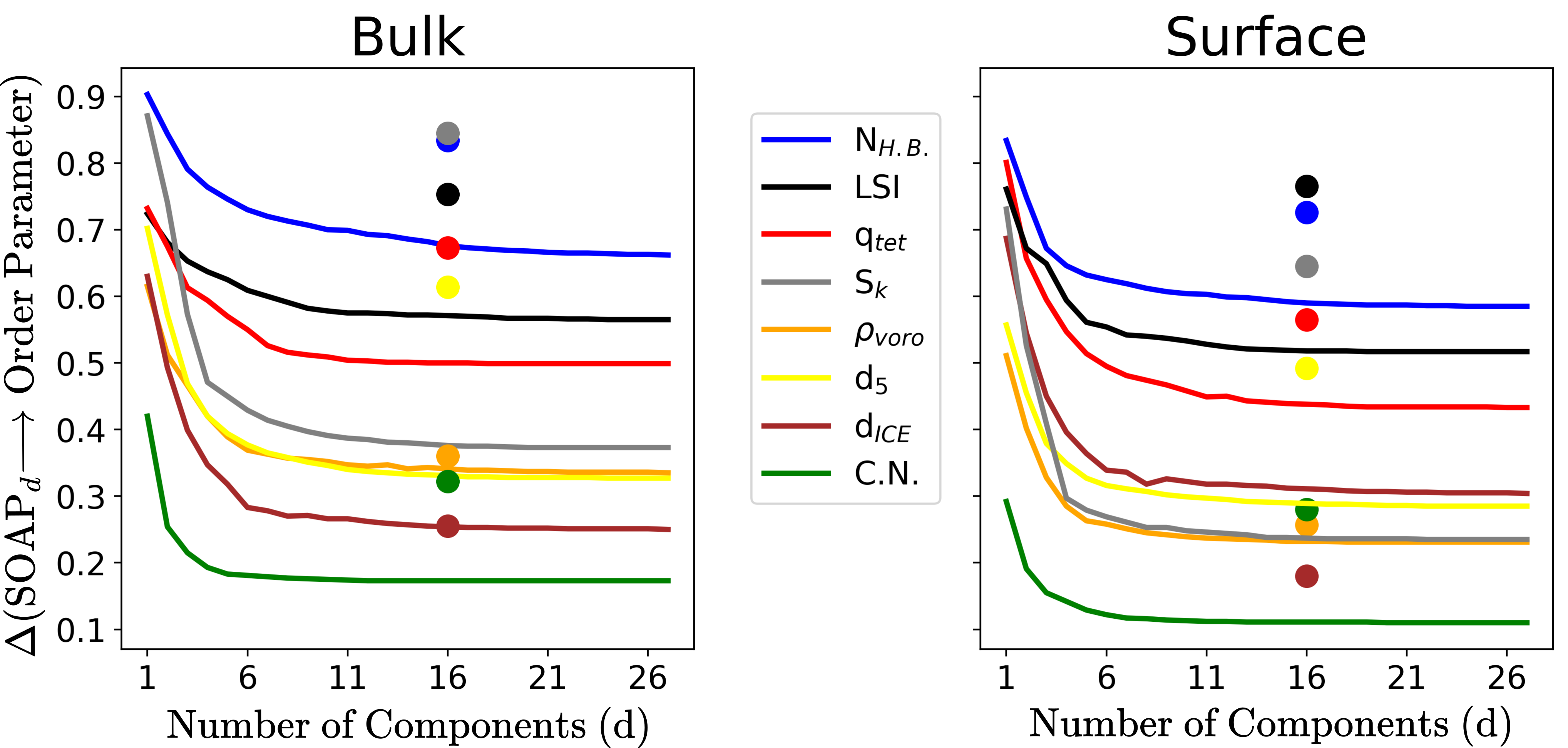}
    \caption{Convergence of the IB between the SOAP space and individual order parameters with $\sigma = 0.25$ \AA \, for both bulk (left) and surface (right)
    . The solid circles represent the optimized IB obtained using $\sigma = 1.0$ \AA \ .
    The rate of convergence of the IB for different variables are different however, with a minimum of about 10-20 SOAP vector components, the minimum IB is reached. 
    Using $\sigma = 0.25$ \AA \ significantly increases the information contained in the SOAP space about the order parameters.
    }
    \label{greedy}
\end{figure}

Figure \ref{greedy} shows the IB between SOAP space and the individual order parameters for both bulk and surface (left and right panels respectively), as a function of the number of components. Note that for the SOAP parameters deployed here, the full power spectrum is a vector with dimension $D=4410$. The easiest variable to predict is the coordination number ($C.N.$) and the most difficult to predict is the total number of hydrogen bonds ($N_{H.B.}$). Overall, the effect of changing from $\sigma=1.0 \ \text{\AA}$ to $\sigma=0.25 \ \text{\AA}$ is consistent with the previous observations seen in Figure \ref{chem_soap} and Figure \ref{select_chem_soap}. The IB is reduced for smaller $\sigma$ - the solid circles shown in both panels correspond to the optimal IB obtained using the larger $\sigma$ value.  By increasing the local atomic resolution, the IB can decrease by up to 50\% and thus enhance the resolution of SOAP for the prediction of  several order parameters. 

In similar spirit to the analysis presented earlier in Figure \ref{select_chem_soap}, the IB between SOAP and order parameters identifies chemical coordinates such as coordination number, $d_{ice}$, $d_{5}$ and $\rho_{voro}$, whose chemical information is well characterized by SOAP. Furthermore, although different order parameters plateau to the optimal IB with different rates, one only needs a total of 10-20 components of the full SOAP power spectrum to predict these order parameters. These trends are found in both bulk and surface liquid water situations.

Perhaps what is more surprising is that SOAP does not seem to be able to predict accurately the number of hydrogen bonds, LSI and $q_{tet}$. On the one hand, this suggests that the SOAP features may not be complete in terms of their information content of the underlying chemical system. However, many of these order parameters are constructed based on strict geometrical cutoffs. For example; along distances or angles or number of neighbours which is then reflected in the IB. 

\begin{figure}[htp]
    \centering
    \includegraphics[width=10cm]{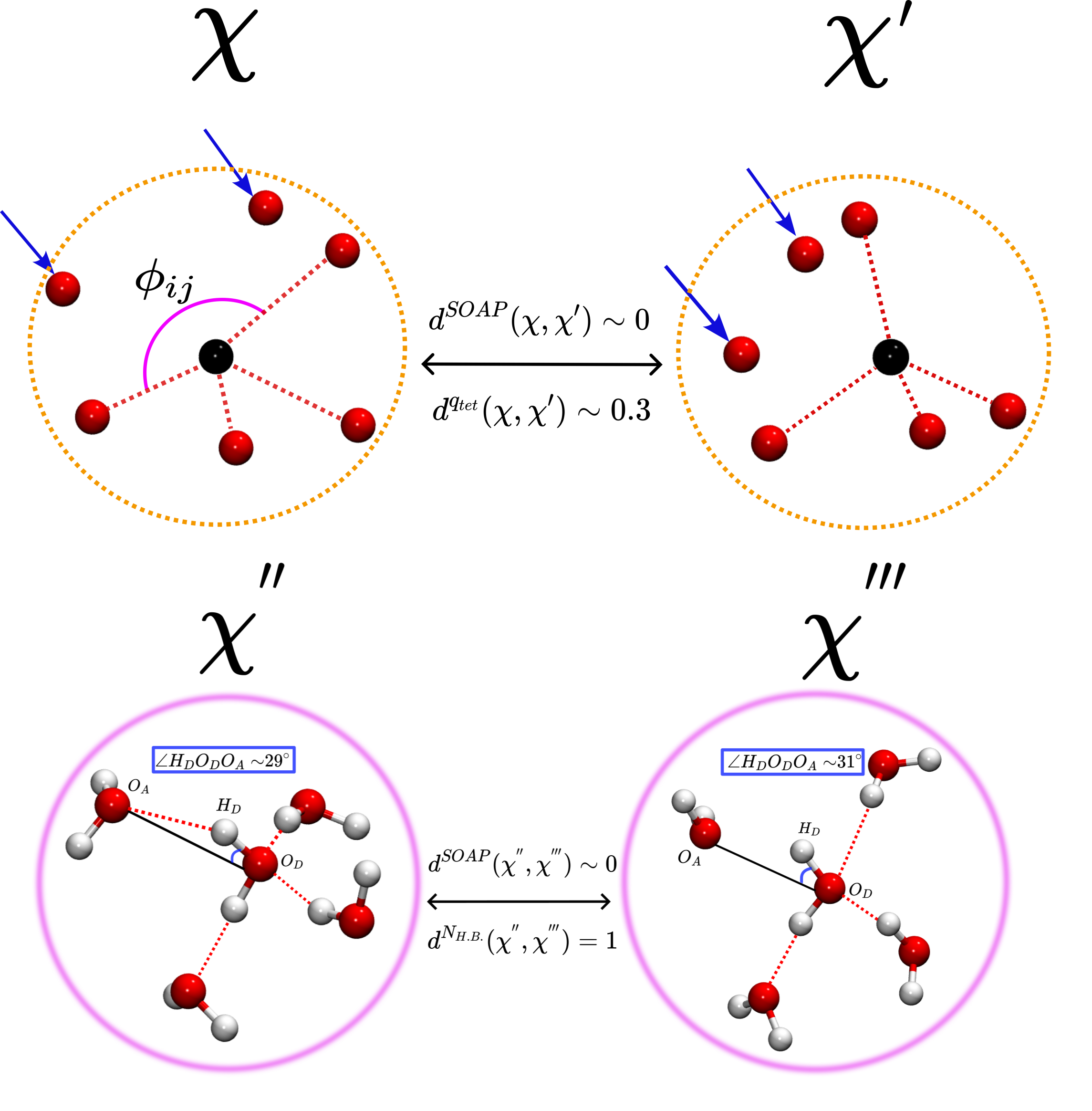}
    \caption{Snapshot of two sets of environments $\chi$, $\chi'$ and $\chi''$, $\chi'''$  which are nearest neighbours in SOAP space but are distant in $q_{tet}$ (top panel) and $N_{H.B.}$ (bottom panel) space. Since the definition of $q_{tet}$ focuses on the first four neighbours of a central atom within the first solvation shell, the two environments are flagged as not tetrahedral (top left) and tetrahedral (top right), while the SOAP space predicts the two environments to be similar.  
    For $N_{H.B.}$, its strict angular cut-off picks up slight angle changes ($\sim$ 2$^{\circ}$) which labels the two environments as having different number of hydrogen bonds. However, these are clearly two environments that are more similar than different.}
    \label{qtete}

\end{figure}

To illustrate this effect more clearly, we show in Figure \ref{qtete} a set of water environments which compares and contrasts distances computed between SOAP, $q_{tet}$ and hydrogen bonding.  We begin with the top panel comparing SOAP and $q_{tet}$. Within an $r_{cut}$ of 3.7 \AA, there are two interstitial oxygen atoms (purple arrow). Since the definition of $q_{tet}$ only looks at the first 4 nearest oxygens, the two environments are flagged as non tetrahedral (left) and highly tetrahedral (right). Due to the restriction of focusing on the nearest 4 oxygens, the $q_{tet}$ parameter picks up an angle that clearly deviates from tetrahedrality. In fact, when one restricts the SOAP computation to only the first four neighbours, the SOAP space is completely predictive of $q_{tet}$ (see SI Figure 10).

In similar spirit to the analysis presented for $q_{tet}$, the bottom panel shows two environments that are again close in SOAP space, but far in terms of hydrogen bonding. Specifically, the bottom left panel shows an environment that accepts and donates two hydrogen bonds (N$_{H.B.} = 4$) while the bottom right shows a small change in the local geometry where the angle used for the hydrogen bonding criterion changes from 29$^{\circ}$ to 31$^{\circ}$ resulting in a defect that now accepts two but donates only one hydrogen bond (N$_{H.B.} = 3$). In hydrogen-bonding space, these configurations are topologically different but in SOAP space, these are, unsurprisingly very similar.

\begin{figure}[htp]
    \centering
    \includegraphics[width=15cm]{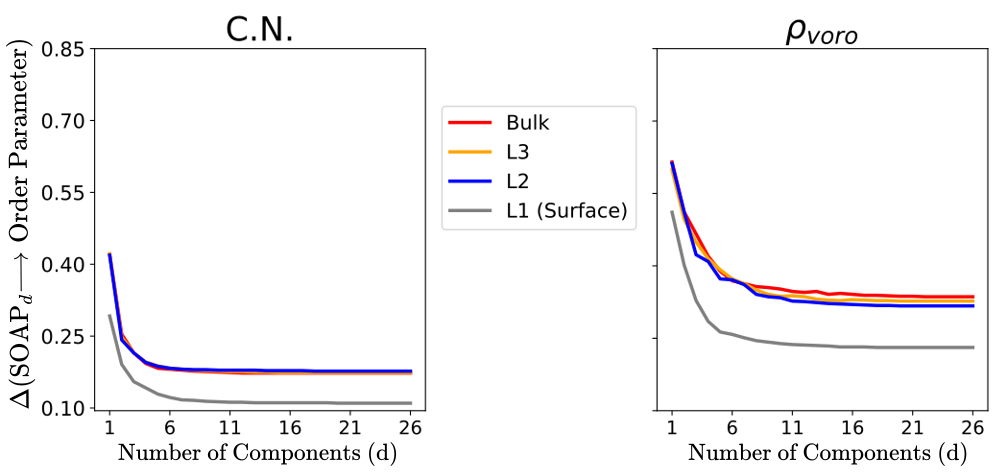}
    \caption{Convergence of the IB between SOAP and selected order parameters, as a function of the number of SOAP vector components. The compression is done for all the layers of the density profile depicted in the supplementary material.
    These results confirm previous reports showing that most of the significant structural changes in the water environments occurs in the topmost layer.}
    \label{layers_soap_chem}
\end{figure}

\subsection{Predictions Within Order Parameters}

Our analysis shows that there is a complex interplay of different order parameters and that they need to be used in combination in order to predict SOAP similarity both in the bulk and at the air-water interface. In the field of aqueous science where the structure and dynamics of liquid water is studied in terms of different local structures, it is quite common to synonymously associate the different order parameters such as $d_{5}$, LSI and $q_{tet}$ . However, these order parameters are not equivalent.


\begin{figure}[!ht]
    \centering
    \includegraphics[width=15cm]{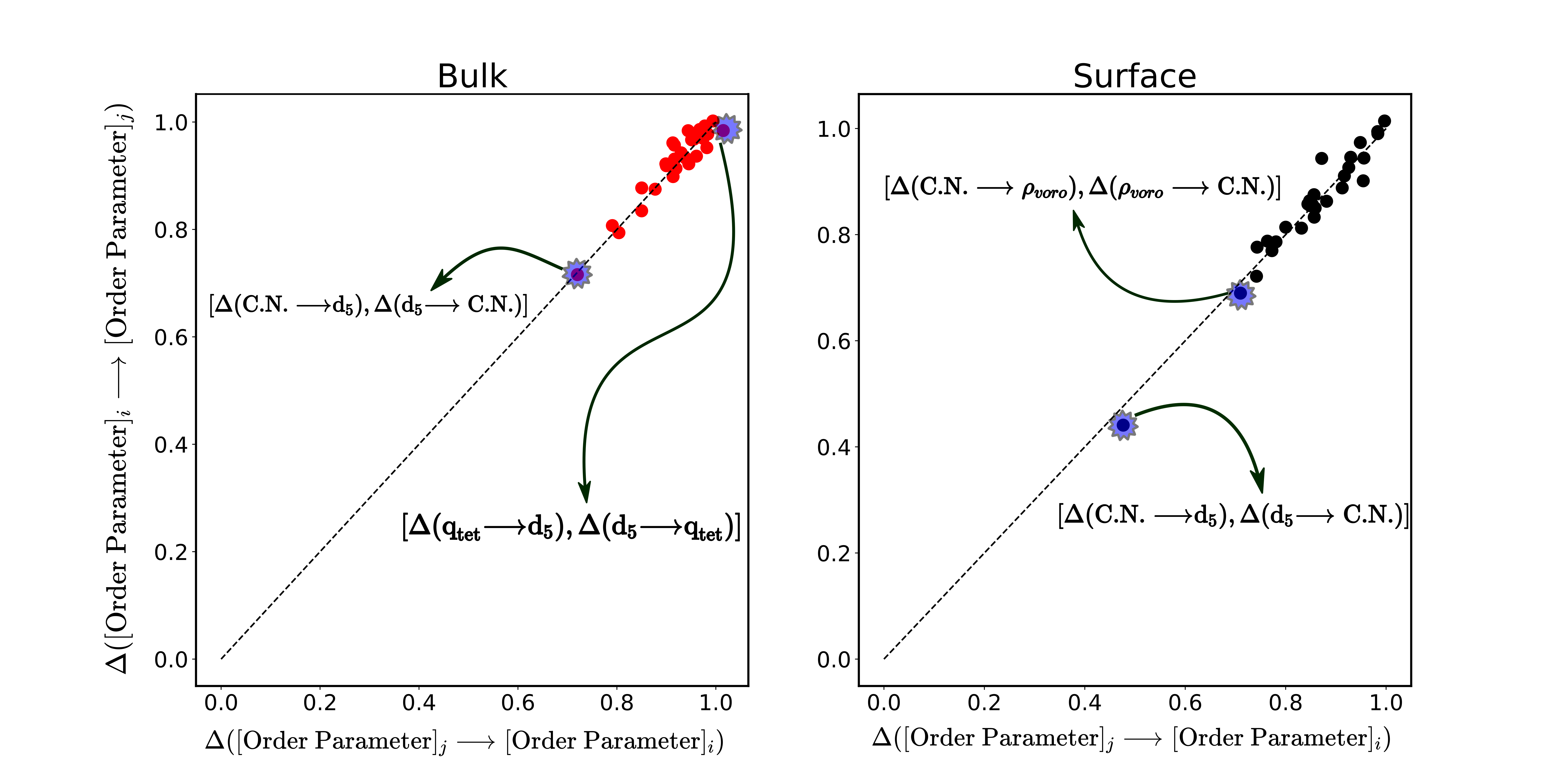}
    \caption{Information Imbalance plane \cite{glielmo2022ranking} for the space of order parameters, showing the IB obtained between different pairs of order parameters. In the Bulk (left) the lowest IB obtained is $\sim$ 0.7 which is between $C.N.$ and $d_5$. The rest of the variables are all not informative of each other.
    In the Surface (right), the lowest IB obtained $\sim$ 0.4 is also between $C.N.$ and $d_5$. In general, most of the variables are uninformative of each other both in the Bulk and on the Surface.}
    \label{ib_plane}
\end{figure}

Figure \ref{ib_plane} shows the IB between all pairs of order parameters for both bulk and surface environments. For bulk environments, we observe that the lowest imbalance obtained is approximately 0.7; between the coordination number and $d_{5}$. Essentially all the other order parameters are not informative of each other consistent with previous work from our group\cite{offei2022high}. Moving to the surface environments marginally improves the predictability for the coordination number and $d_{5}$ pair. All in all, it is clear that the different order parameters used to characterized local water structure contain very little information about each other. For example, considering the IB between tetrahedrality (q$_{tet}$) and local density ($\rho_{voro}$) which is closer to 1, shows that they provide different information about the local environment. Thus, an environment that is more tetrahedral may not have a lower local density and vice-versa. 

\subsection{Diagnostics for Water Structure}

We have used the IB approach to examine the coupling between ML-based local atomic descriptors and chemistry-inspired order parameters. We then explored whether this could be used as a diagnostic tool to study the length-scale of the perturbations induced by the prescence of the surface.

Earlier, we compared the IB for bulk and surface water environments, where the latter focused on interfacial waters that reside only within $\sim$ 3.0 \AA \ of the WCI. Examining the relative density of water with respect to the WCI (see SI Figure 1), one observes that the correlations albeit weaker, extend up to 1 nanometer from the surface. The extent of the thickness of this interface as probed by both surface sensitive vibrational spectroscopy experiments and simulations, and how it is manifested in terms of local structure, continues to be a topic of active research. \cite{pezzotti20172d,zhang2011ultrafast,hsieh2014aqueous}.

Since the IB appears to provide a sensitive measure of the surface water environments, we performed the compression of the full SOAP space for selected order parameters separately for different layers of water with respect to the WCI (see SI Figure 1 for an illustration of how the layers are defined). The left and right panels of Figure \ref{layers_soap_chem} shows the IB between SOAP and the coordination number and $\rho_{voro}$ respectively, as a function of the number of SOAP components. Our results show that the IB in the second and third layer are essentially indistinguishable from the bulk. This shows that there is essentially only a very thin layer of water ($\sim$ 3.0 \AA \ near the surface) whose local structural properties are significantly different, consistent with previous studies \cite{pezzotti20172d}.

\section{Conclusions}
\label{conclusion}

In this work, we have used a recently developed statistical test (IB) to examine the relationships between the SOAP descriptors and chemically inspired order parameters that are used to characterize local water structure. We focus our analysis on how these relationships evolve as one moves from bulk water to the hydrophobic surface of water.

Examining the IB between the full SOAP space and order parameters shows that these two classes of variables do not contain the same information. While the chemical information encoded in the SOAP descriptors can be improved by reducing the Gaussian smearing, the IB between the order parameters and the SOAP descriptors obtained is far from optimal. The predictability improves marginally for water environments sampled at the interface. Given that the SOAP descriptors are designed to be generic and do not probe specific chemical details, this observation may not be too surprising. The consistency between SOAP and the order parameters could be improved by including features that probe directly the fluctuations of the hydrogen bond network. 

Order parameters like the ones we have investigated here to probe the local structure of water, are used ubiquitously for analyzing molecular dynamics simulations of liquid water. In particular, these are often used to assign whether the water structure arises from a low or high density local environment\cite{hamm2016markov,appignanesi2009evidence,offei2022high}. However, our analysis suggests that most of these order parameters are somewhat independent of each other and cannot be used synonymously. Furthermore, the SOAP descriptors seem to be probing the local environment of water molecules in a much less specific manner. One may argue that order parameters such as the hydrogen-bonding criterion are designed to be very sensitive to subtle changes in local topology which cannot be captured by SOAP. We believe that our results suggest that sensitivity analysis of geometric criteria used in chemistry-inspired order parameters would be appropriate.

\bibliography{biblio}

\renewcommand{\thefigure}{S\arabic{figure}}
\section{Supplementary Information}
\setcounter{figure}{0}    
\begin{figure}[!htp]
    \centering
    \includegraphics[width=14cm]{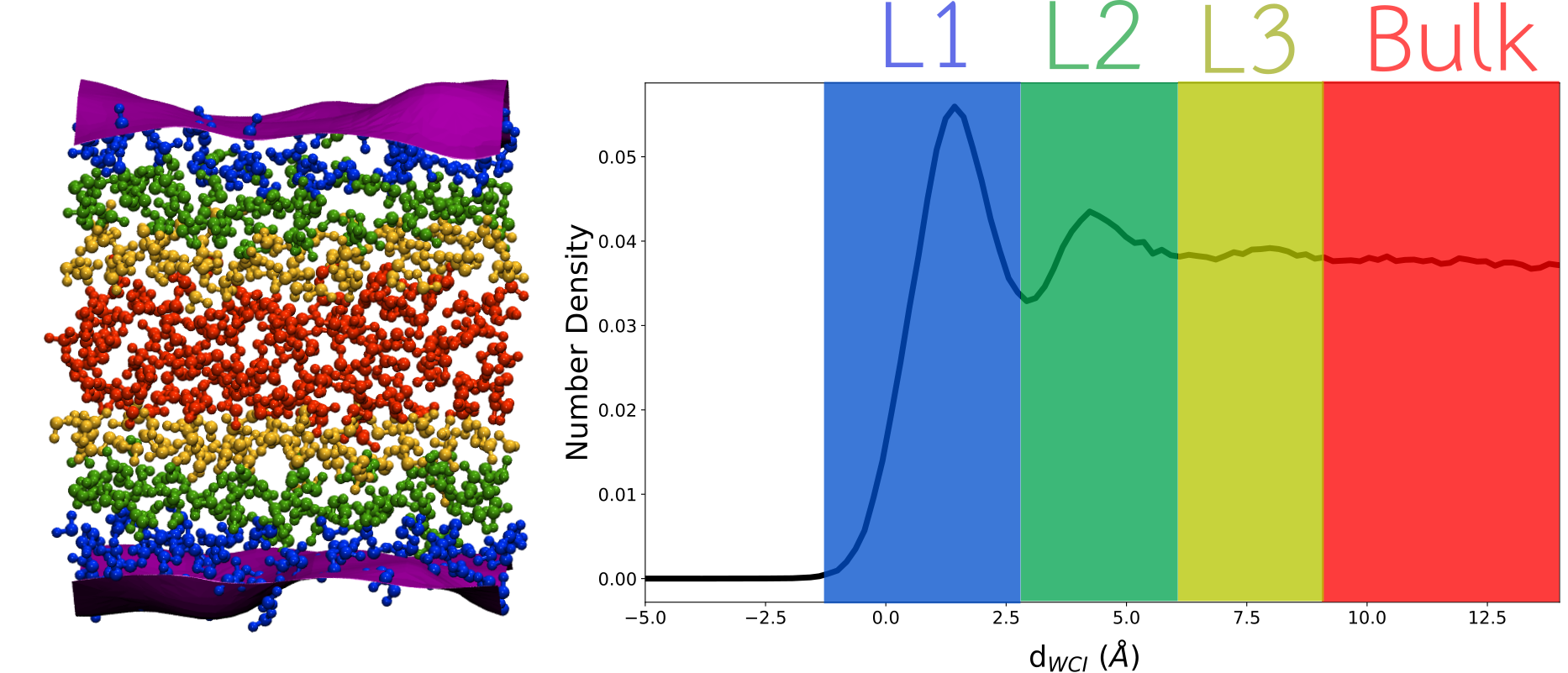}   
    \caption{Snapshot of MD simulaton box (Left)-The water molecules are color-coded based on their distance from the WCI.
    Number density profile of water molecules as a function of the distance from the WCI (Right), the colors correspond to the distinct structural layers that occur and have been labelled (L1-L4) for Surface-Bulk respectively.}
\end{figure}

\begin{figure}[htp]
    \centering
    \includegraphics[width=14cm]{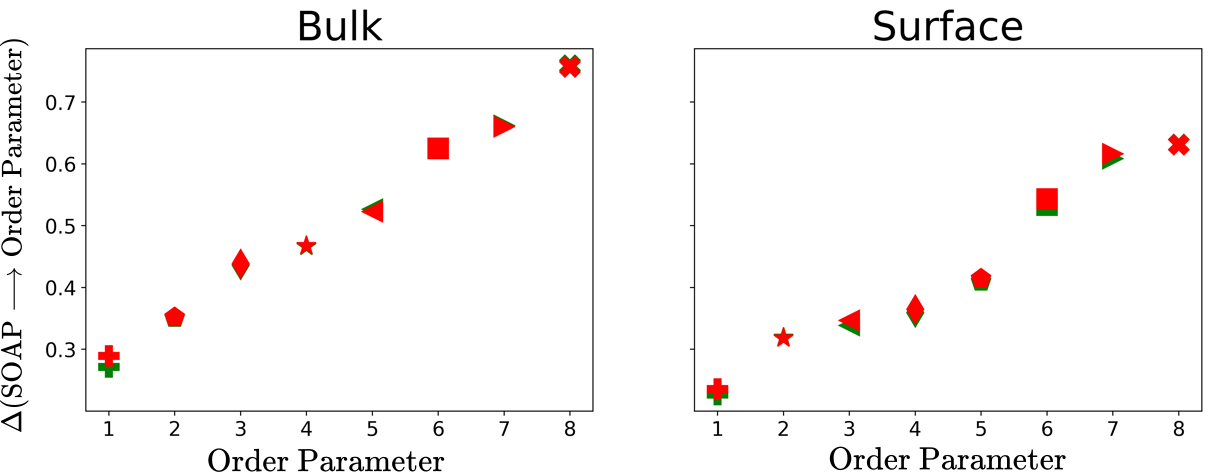}
    \caption{$\Delta (\text{SOAP} \longrightarrow \text{Order Parameters})$ for $\sigma = 0.25$ using ($n_{max}$,$l_{max}$) = (10,8) (Red Symbols) and ($n_{max}$,$l_{max}$) = (10,6) (Green Symbols). For fixed $n_{max}$, the predictability does not change with an increase in $l_{max}$ }
\end{figure}

\begin{figure}[htp]
    \centering
    \includegraphics[width=14cm]{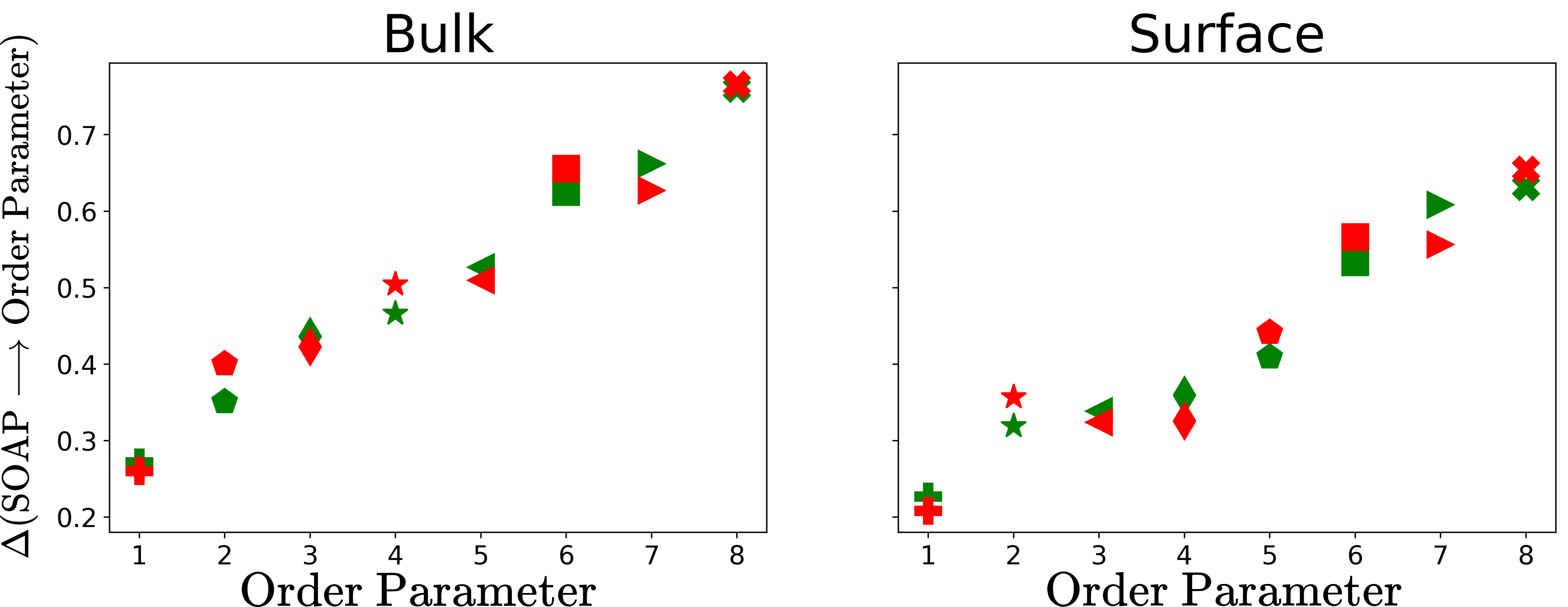}
    \caption{$\Delta (\text{SOAP} \longrightarrow \text{Order Parameters})$ for $\sigma = 0.25$ using ($n_{max}$,$l_{max}$) = (12,6) (Red Symbols) and ($n_{max}$,$l_{max}$) = (10,6) (Green Symbols). For fixed $l_{max}$, the predictability does not change with an increase in $n_{max}$ }
\end{figure}

\begin{figure}[htp]
    \centering
    \includegraphics[width=14cm]{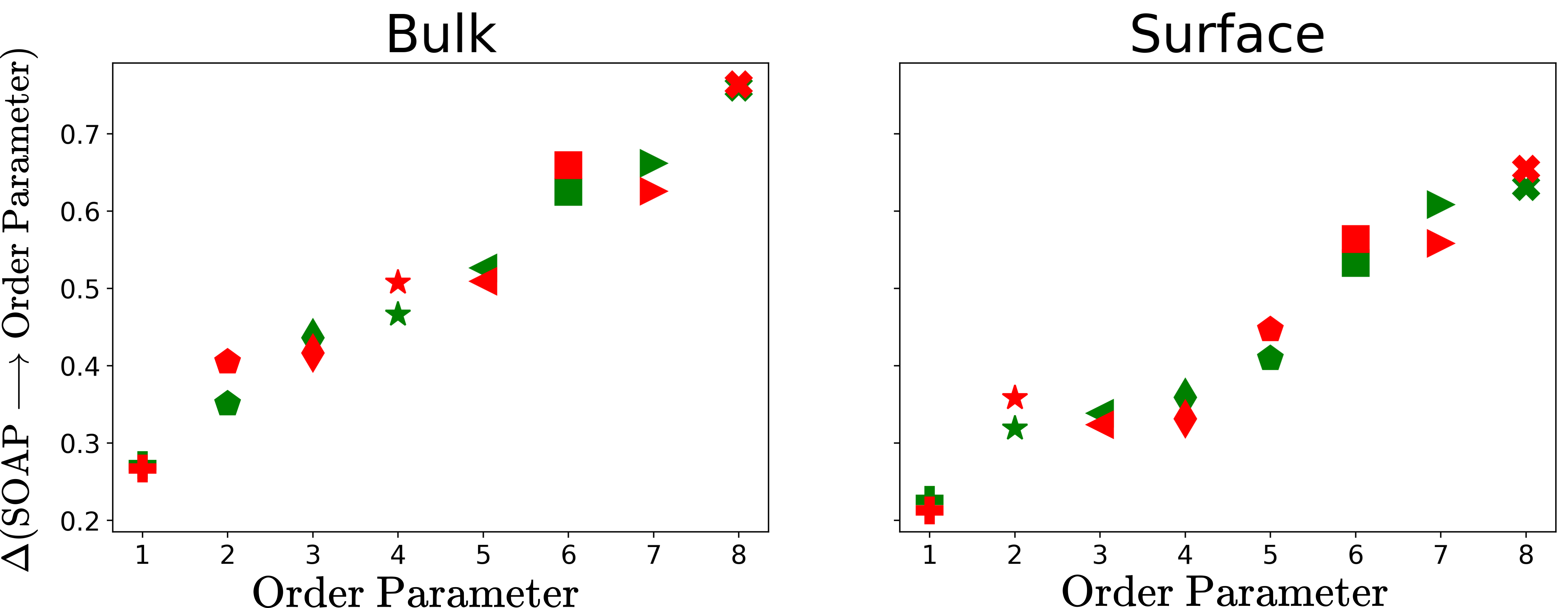}
    \caption{$\Delta (\text{SOAP} \longrightarrow \text{Order Parameters})$ for $\sigma = 0.25$ using ($n_{max}$,$l_{max}$) = (12,8) (Red Symbols) and ($n_{max}$,$l_{max}$) = (10,6) (Green Symbols). Changing both $n_{max}$ and $l_{max}$ does not significantly change the predictabilities}
\end{figure}

\begin{figure}[htp]
    \centering
    \includegraphics[width=14cm]{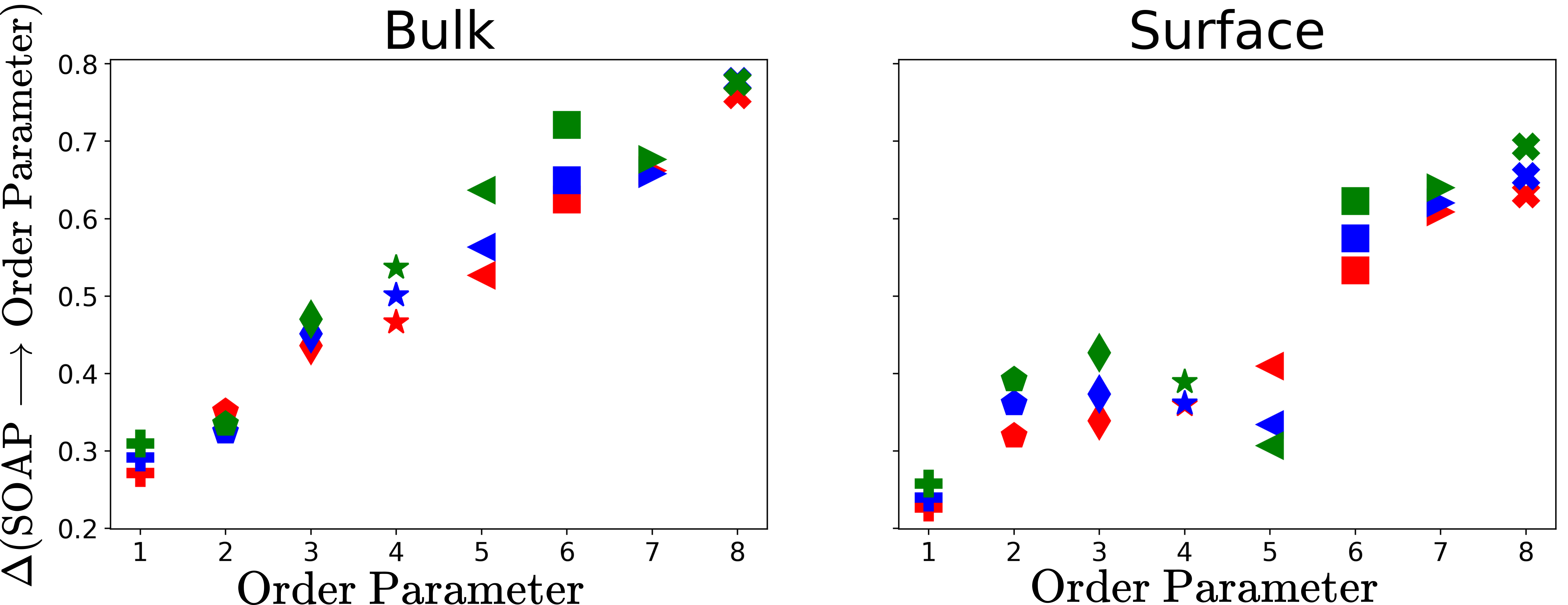}
    \caption{$\Delta (\text{SOAP} \longrightarrow \text{Order Parameters})$ for $\sigma = 0.25$ and $r_{cut}$ = 3.7 (Red), $r_{cut}$ = 4.5 (Blue) and $r_{cut}$ = 5.5 (Green). In general, an $r_{cut}$ of 5.5 makes the SOAP consistently less informative about the chemical variables (although very slightly). The main thing to notice is that an $r_{cut}$ of 3.7 is able to capture as much as SOAP can, all the local fluctuations contained in the Order Parameter.}
\end{figure}

\begin{figure}[htp]
    \centering
    \includegraphics[width=15cm]{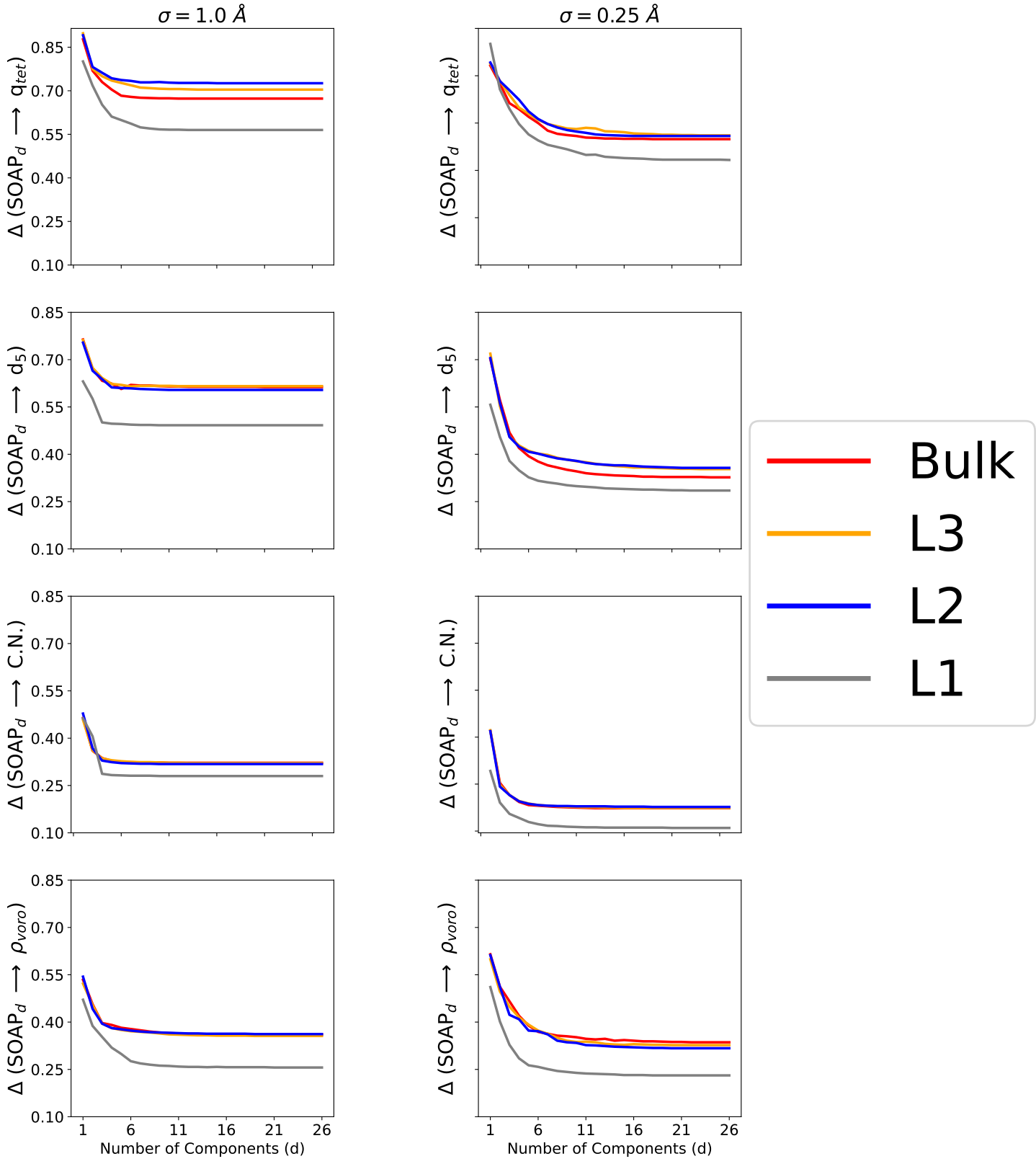}
    \caption{Evolution of $\Delta (\text{SOAP} \longrightarrow \text{Order Parameters})$ as a function of the layer definitions showed in SI figure 1 for $\sigma = 1.0$ \AA \ (Left) and $\sigma = 0.25$ \AA \ (Right). There is a general decrease in the IB as we go from a larger to a smaller $\sigma$.
    This also confirms that much of the structural changes that occur is at the topmost layer (L1)}
\end{figure}

\begin{figure}[htp]
    \centering
    \includegraphics[width=15cm]{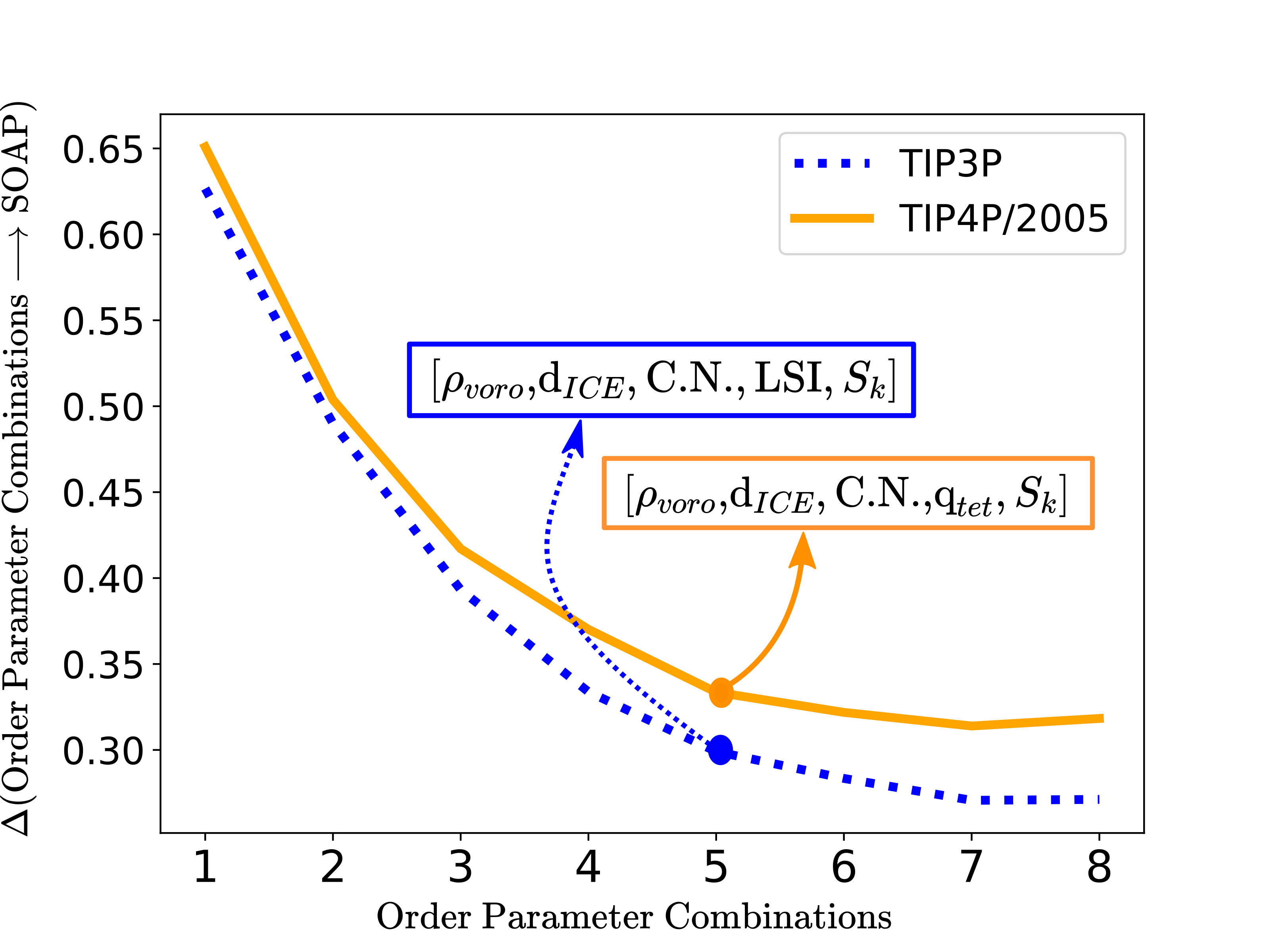}
    \caption{$\Delta (\text{Order Parameters} \longrightarrow \text{SOAP}$) for the TIP3P and TIP4P/2005 water model for the Surface layer. The optimum set of 5-variables selected are consistent across the two models with only a difference in one selected variable ($\text{q}_{tet}$ for the TIP4P/2005 and $\text{S}_k$ for TIP3P).}
\end{figure}

\begin{figure}[htp]
    \centering
    \includegraphics[width=15cm]{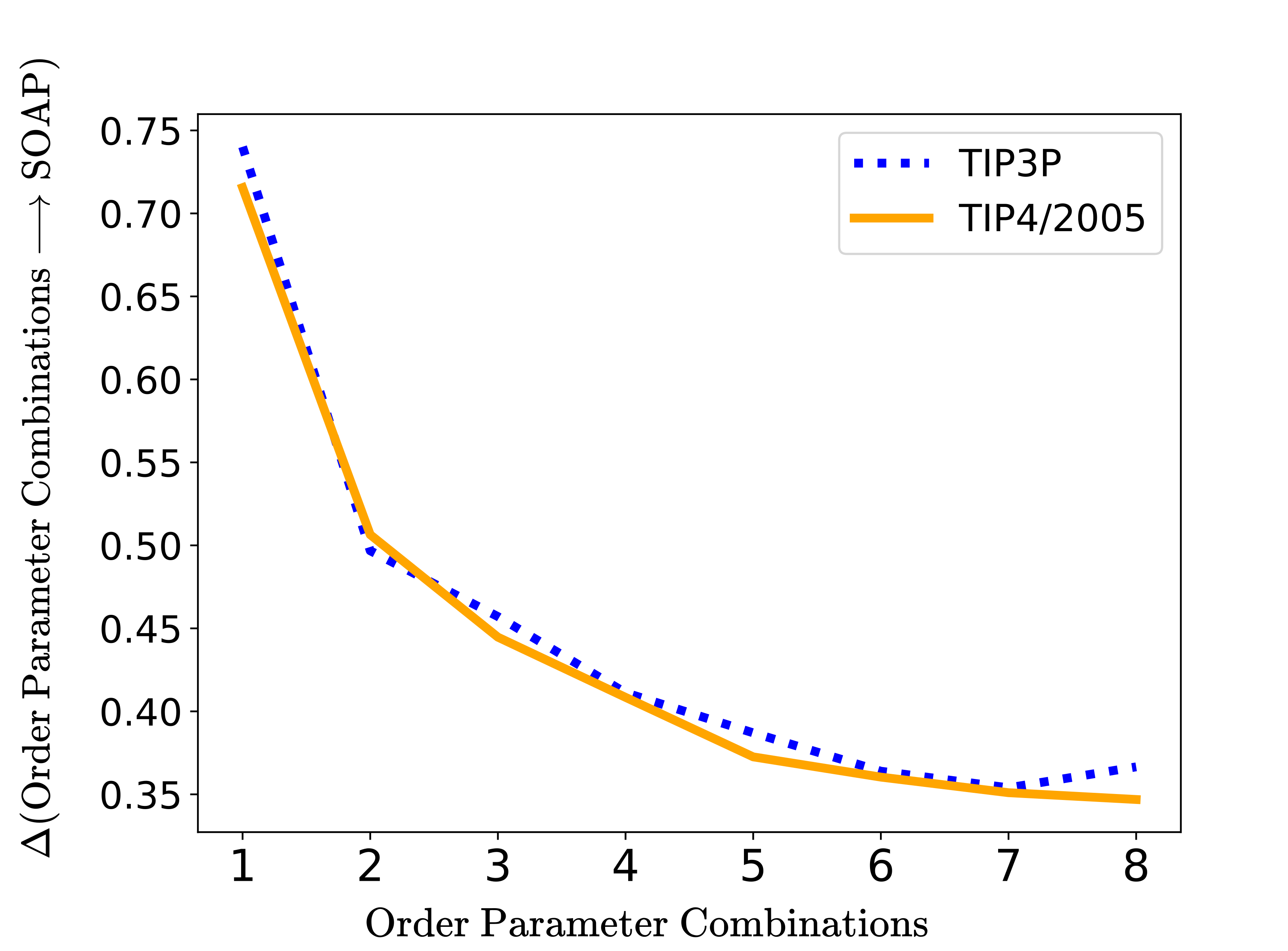}
    \caption{$\Delta (\text{Order Parameters} \longrightarrow \text{SOAP}$) for the TIP3P and TIP4P/2005 water model for the Bulk layer. The results are consistent across the two water models.}
\end{figure}

\begin{figure}[htp]
    \centering
    \includegraphics[width=15cm]{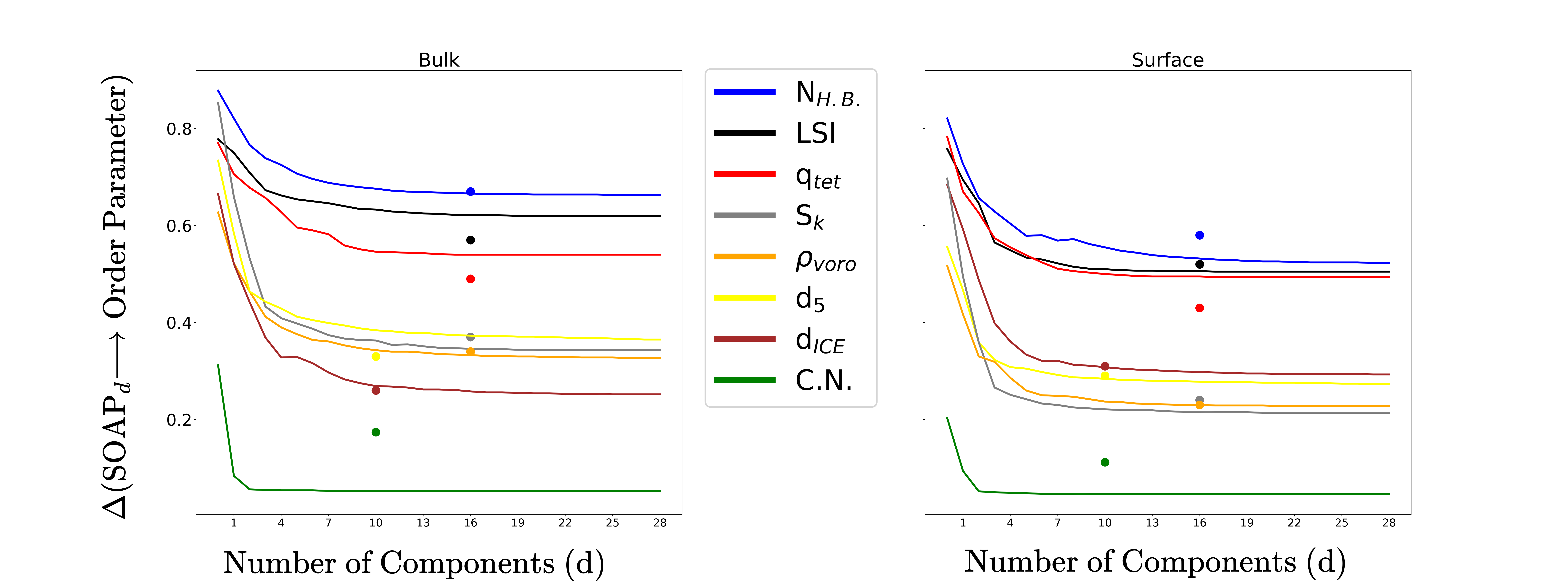}
    \caption{Convergence of $\Delta (\text{SOAP} \longrightarrow \text{Order Parameters})$ as a function of the number of SOAP components for the TIP3P water model. The solid circles represent the optimized IB obtained for the TIP4P/2005 model for the specified color coded chemical variable. These results show very small differences in the IB across the two models and hence are consistent.}
\end{figure}

\begin{figure}[htp]
    \centering
    \includegraphics[width=15cm]{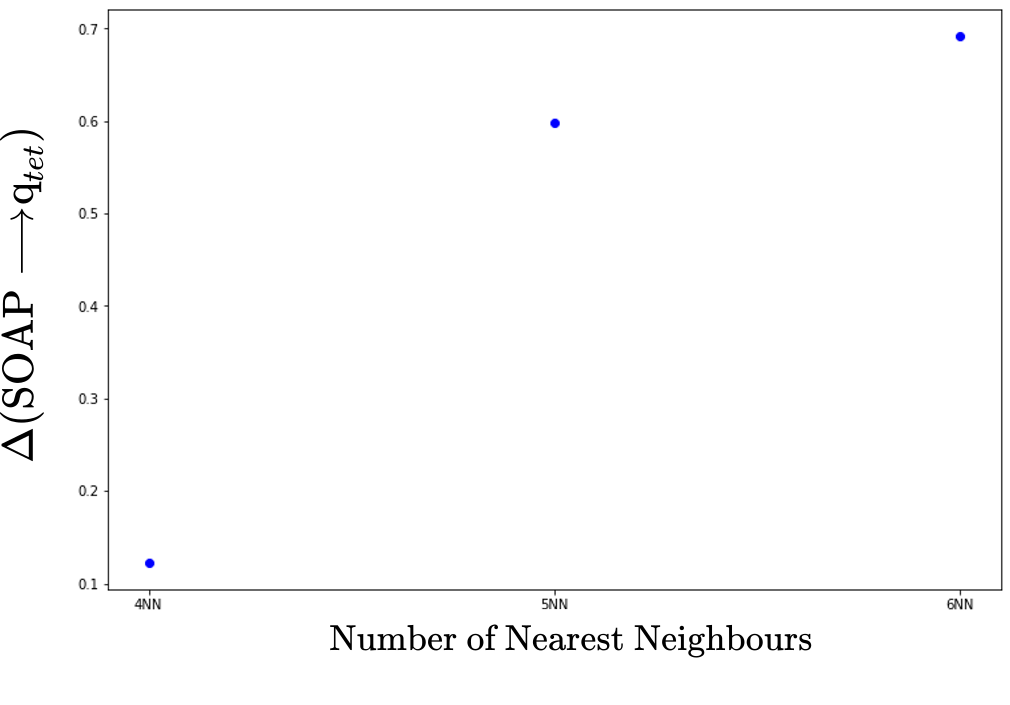}
    \caption{Information Imbalance between SOAP and $\text{q}_{tet}$ for SOAP descriptors computed with 6,5 and 4 nearest neighbours. As the number of neighbours used to compute the SOAP descriptors decreases, the ability of SOAP to predict $\text{q}_{tet}$ increases. 
    With 4 nearest neighbours, the SOAP space is completely predictive of $\text{q}_{tet}$}
\end{figure}

\end{document}